%% file: tims-in-practice.tex
\documentclass{IEEEcsmag}

\usepackage[colorlinks,urlcolor=blue,linkcolor=blue,citecolor=blue]{hyperref}

\usepackage{upmath}
\usepackage{booktabs}

\usepackage{todonotes}
\usepackage{comment}

\usepackage{multirow}

\usepackage[most]{tcolorbox}

\definecolor{algborder}{cmyk}{0.85,0.38,0,0} 
\definecolor{algtitlebgcolor}{cmyk}{0,0,0,0.80}

\newtcolorbox{boxone}[1][]{enhanced, breakable,rounded corners=southeast,
toptitle=6pt,bottomtitle=7pt,title=#1,
attach boxed title to top,
          boxed title style={colframe=black},
titlerule style={black,line width=1pt}, 
coltitle=white,colbacktitle=algtitlebgcolor, 
before skip=10pt,boxrule=1pt,boxsep=0pt,left=12pt,right=12pt,top=6.5pt,
bottom=12pt,arc=8pt
}
\newcommand{\detailtexcount}[1]{%
	\immediate\write18{texcount -merge -sum -q #1.tex > #1.wcdetail }%
	\verbatiminput{#1.wcdetail}%
}

\usepackage{capt-of}

\jvol{XX}
\jnum{XX}
\paper{8}
\jmonth{May/June}
\jname{IEEE Software}
\pubyear{2020}

\input{macro-editing}

\begin{document}

\sptitle{Department: Head}
\editor{Editor: Name, xxxx@email}

\title{Managing Traceability Information Models: Not such a simple task after all?}

\author{Salome Maro, Jan-Philipp Stegh\"ofer, 
Eric Knauss, Jennifer Horkoff, Rashidah Kasauli}
\affil{Chalmers $\mid$ University of Gothenburg, Sweden}

\author{Rebekka Wohlrab}
\affil{Systemite AB and Chalmers $\mid$ University of Gothenburg, Sweden}

\author{Jesper Lysemose Korsgaard, Florian Wartenberg, Niels J\o{}rgen Str\o{}m}
\affil{Grundfos, Denmark}

\author{Ruben Alexandersson}
\affil{Volvo Cars, Sweden}

\markboth{Department Head}{Paper title}

\begin{abstract}
Practitioners are poorly supported by the scientific literature when managing traceability information models (TIMs), which capture the structure and semantics of trace links. 
In practice, companies manage their TIMs in very different ways, even in cases where companies share many similarities. 
We present our findings from an in-depth focus group about TIM management with three different systems engineering companies. We find that the concrete needs of the companies as well as challenges such as scale and workflow integration are not considered by existing scientific work. We thus issue a call-to-arms for the requirements engineering and software and systems traceability communities, the two main communities for traceability research, to refocus their work on these practical problems.
\end{abstract}

\maketitle

\chapterinitial{A crucial building block}
in a successful traceability strategy for any organisation is a well-defined and goal-oriented traceability information model (TIM). A TIM defines the relationships between the diverse artifacts created during system development, often by different teams and often different parts of an organisation. For these relationships, it defines structural aspects such as trace links types and cardinalities. Additionally, a TIM's semantic information can be informed by organisational structures, information about the workflow, the interaction between products and platforms, and the collaboration between different parts of an organisation. This makes a TIM a crucial artifact in an organisation and, at the same time, very difficult to change once established. The main question when defining a TIM is thus how to capture these aspects and translate them into a model that the individual engineer can successfully work with and use to create value. TIM management includes activities such as defining which relationship types are needed, implementing the designed TIM in a tool and managing its evolution. 
Because of the complexity of what a TIM captures and its importance for traceability, each company, and sometimes each project, requires a tailored TIM.


Due to this specificity, practitioners lack concrete guidelines 
when creating TIMs especially in large systems engineering companies. To address this issue, we conducted a workshop with three companies  developing  systems from  three  different  domains to understand their current and emerging TIMs and the challenges they face when managing them.  The  workshop  revealed  that each of the three companies manage their TIMs in radically different ways but that they face common challenges which are yet to be addressed in literature. We summarise our findings in this paper and describe research challenges that result from them as a call-to-arms for the RE and software and systems traceability community.
\begin{boxone}[{{Important terms}}]
\textbf{Traceability} is is defined as the ability to interrelate any uniquely identifiable software engineering artifacts to any other, maintain required links over time, and use the resulting network to answer questions of both the software product and its development process~\cite{COEST}. 

\textbf{A traceability strategy }is a plan of action 
leveraged to design a traceability solution, consisting of a traceability process, tooling, and a traceability information model (TIM).

\textbf{A traceability information model (TIM)} defines all the artifact types that need to be traced to and the different relationship types between the artifacts types. Figure~\ref{fig:tim-example} shows an example of a simple TIM. 
	\includegraphics[width=\textwidth]{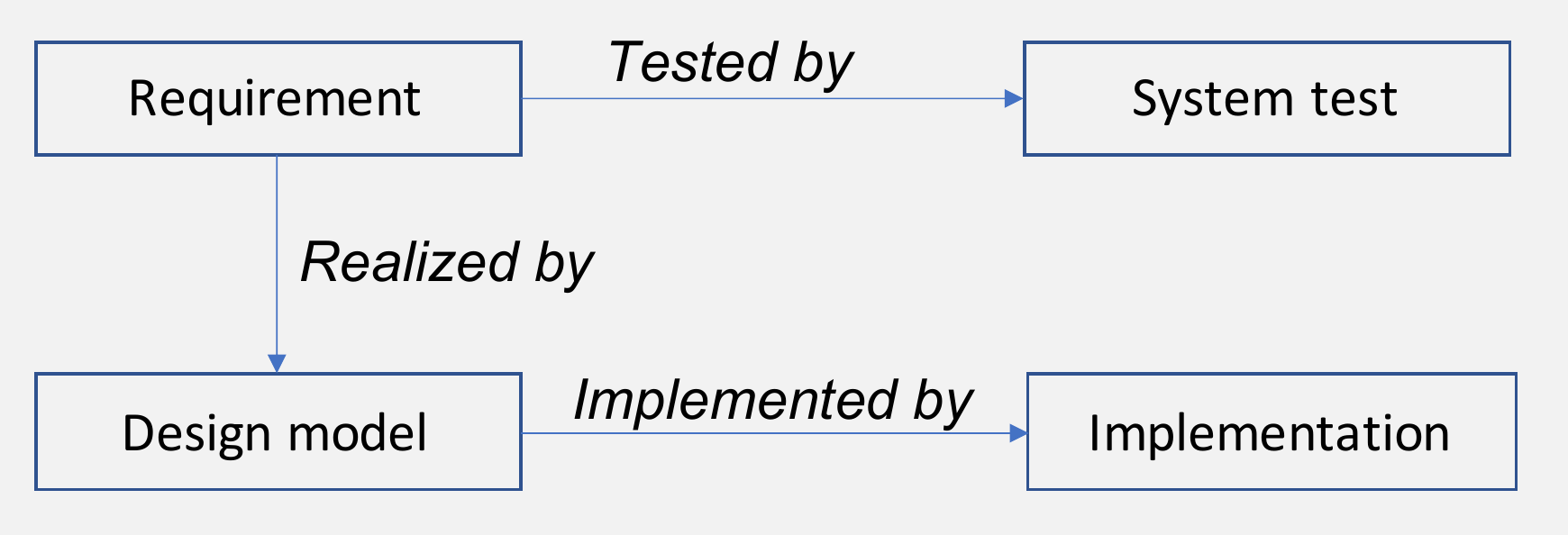}
	\captionof{figure}{A simple TIM as an example. It connects four different artefact types (requirements, system tests, design models, and implementation) with three different trace link types (tested by, realized by, and implemented by).}
	\label{fig:tim-example}
\end{boxone} 
 
\begin{table*}[t]
\centering
\caption{Overview of the three companies that participated in the workshop. In order to protect each company's anonymity, the text refers to Company 1, 2, and 3 instead.}
\label{table:three_companies}
\begin{tabular}{@{}lp{3.9cm}p{3cm}p{4.8cm}@{}}
\toprule
 & \textbf{Company A}          & \textbf{Company B}        & \textbf{Company C} \\
\midrule
Domain & Industrial automation &  Telecommunication & Automotive  \\
No.\ of participants &  3 &  1 &  3\\ 
Participants' roles & 2 Lead Systems Engineers\newline 1 System Test Engineer   &  1 Systems Engineer & 1 Information Architect\newline 2 Process, method, and tools specialists\\ 
\bottomrule
\end{tabular}
\end{table*}

\section{Background on TIM management}
The literature on TIMs management is split into three main themes: research proposing TIMs for specific purposes e.g., TIMs for safety certification~\cite{cleland2012trace} or for managing non-functional requirements~\cite{kassab2008traceability}; research proposing generic TIMs that can be reused, e.g.,~\cite{espinoza2005need}; and finally tools that support concrete implementation of TIMs that have already been designed, such as~\cite{drivalos2008engineering}.
Regardless of theme, however, many of the studies are abstract (e.g,~\cite{ramesh2001toward}) or conducted using small examples that are isolated from realistic company environments. Their generalisability is thus dubious. In both the research community and in industry, it is agreed that no single TIM fits every  development context, therefore, the number of studies proposing generic TIMs has decreased over time, and research is focused on project-specific or task-specific TIMs~\cite{cleland2014software}.


While it is true that no single TIM fits every development context, concrete guidelines are needed to support practitioners to define, implement and maintain their specific TIMs. This part of research is lacking and practitioners have to manage TIMs with little support from scientific literature. A few studies exist, e.g., the work by Ramesh and Jarke~\cite{ramesh2001toward} provides some reference TIMs that practitioners could use as a starting point for designing their own TIMs. The paper discusses issues encountered when using the reference models in few case studies.  Additionally, M\"ader et al.~\cite{mader2009getting} gives a few guidelines for defining TIMs and provides guidelines for traceability in safety-critical environments which also include some guidelines on TIM design~\cite{mader2013strategic}. Overall, different aspects are still missing such as as alignment of TIMs with different processes and workflows, TIM evolution and in general TIM management in large organisations. 

Our long-term goal is to fill this gap. This paper presents the first step in this endeavour, i.e., identifying practitioner's challenges when managing TIMs, in order to synthesize useful solutions and guidelines for these challenges.

\begin{boxone}[{{Methodology}}]
We conducted a focus group with five researchers and seven practitioners from three companies from the systems engineering domain. 
All participants from the companies have experience with traceability. 
Additionally, participants from two of the three companies 
have practical experience designing and managing TIMs.
During the focus group, one of the researchers presented examples of TIMs and a summary of guidelines on how to define TIMs extracted from literature. This was followed by presentations from the three companies (Table~\ref{table:three_companies}) on their TIM and their TIM management process.  Each presentation was followed by a discussion of the specific challenges presented and the relationship to literature. The discussion was open such that all participants could  ask questions about the presented TIM and TIM management process. The researchers took notes during the presentation and discussions. 

The notes were later analyzed in a brainstorming session with five researchers. The challenges identified were sorted and grouped into three categories and member checked with the industry participants of the focus group. 

In order to protect each company's anonymity, we refer to Company 1, 2, and 3 instead, when reporting results. These numbers are not in any particular order; specifically, Company 1 does not necessarily correspond to Company A.

\end{boxone} 

\section{Practitioners' Challenges}
\label{sec:challenges}

The three companies that presented their TIMs in the workshop all have a different approach towards the design of TIMs. 
 In \emph{Company 1}, the TIM is 
not formally documented as a model, but implicitly defined as part of the definition of done and integrated in the continuous integration tooling. 
Developers are thus aware of which trace link types to create. 
The TIM focuses on linking development artifacts in an agile environment, e.g., test cases, user stories and code. \emph{Company 2} 
defines an overall information model that models all artifacts and aspects used in the development environment. 
The traceability strategy is derived from this information model. 
\emph{Company 3} follows a formal process and defines a company-wide TIM with a centralized approach via a feature model and including traceability in two dimensions: traceability between platform artifacts and between the product instances of the platform artifacts. 

While the companies follow different approaches,  the challenges can be ascribed to three main drivers:
\begin{itemize}
    \item The companies develop \emph{complex products} using a \emph{product line approach}.
    \item The companies are \emph{large and distributed} and different teams need to work together on the same end product.
    \item The companies have \emph{long-lasting products} which means traceability needs to be maintained over a long period of time.
\end{itemize}



\begin{figure*}
    \centering
    \includegraphics[width=\textwidth]{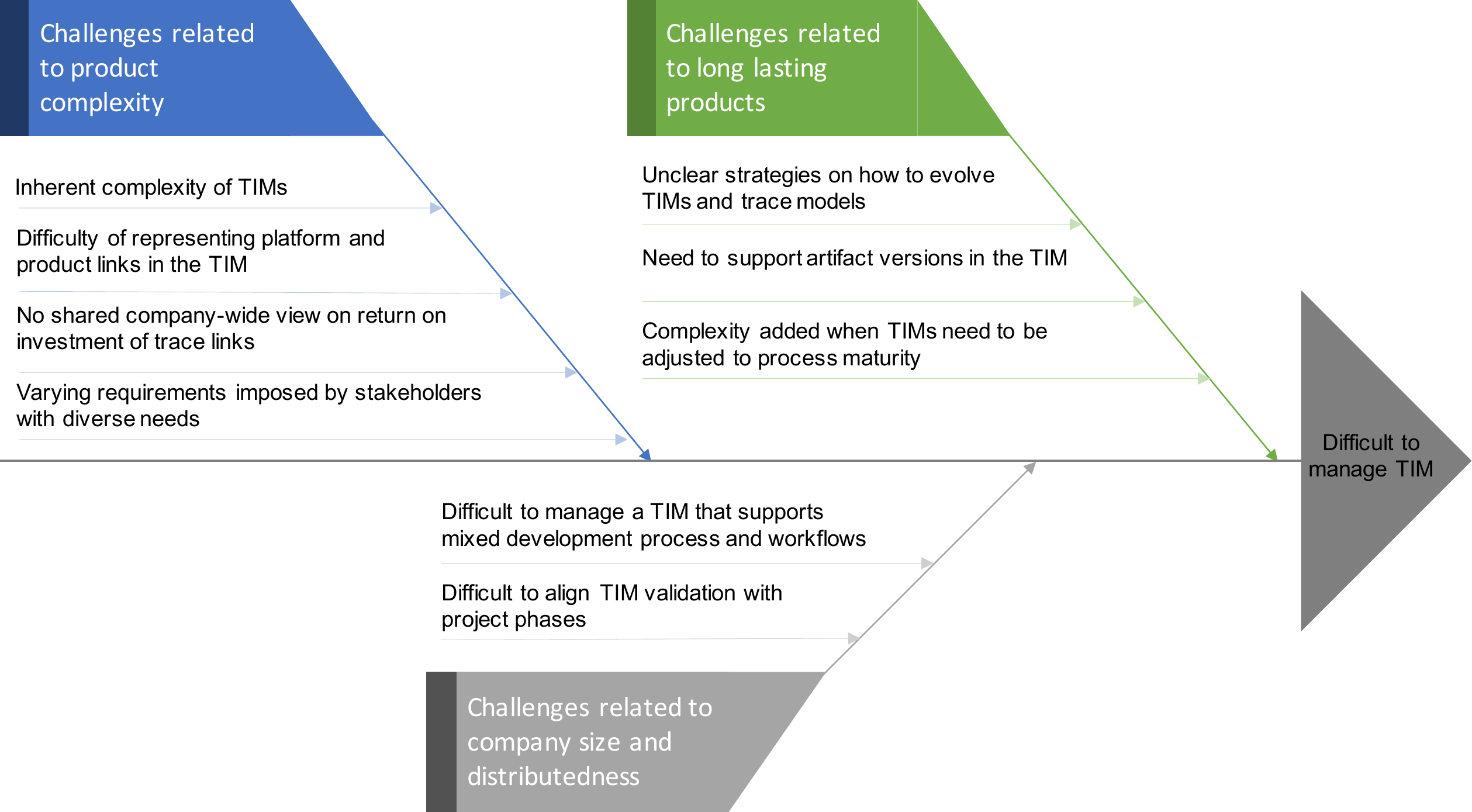}
    \caption{Overview of Challenges that make design of TIMs difficult in practice. 
    }
    \label{fig:tim-chlgs-ishakawa-diagram}
\end{figure*}

The challenges are presented in the sections below and summarised in Figure~\ref{fig:tim-chlgs-ishakawa-diagram}. 

\section{Complex products}
The focus group allowed us to identify challenges that arise due to the complexity of the products developed by the companies. 
\begin{itemize}
\item \textbf{Inherent complexity of TIMs}
A TIM for a large and complex system containing a large number of artifact types needs to encompass the entire product and is therefore complex itself with a large number of artifacts types and relationship types. Thus managing such a TIM is difficult.

\emph{Company 2} addresses this challenge by creating an overall information model. It not only defines all artifact types that exist along with their relationship types, but also captures additional information such as tool capabilities and role responsibilities. While one might argue that the information model constitutes a TIM, it contains aspects that are not relevant for traceability of software development artifacts which are not clearly delineated. For instance, the information model used by \emph{Company 2} contains more than 50 artifact types and also defines workflows and responsible roles. This makes it difficult to reason about traceability with such information models.  

One of the engineers from \emph{Company 2} reported that the complexity of the overall information model makes it difficult to define a subset of the information model as a TIM and plan an overall traceability strategy for the company based on this.
   
\item \textbf{Difficulty of representing platform and product links in the TIM}
In a product line environment, traceability needs to support maintenance of the shared assets in the platform as well as the assets for specific products. In such a scenario, companies struggle with defining a TIM in a way that caters to product lines activities (e.g., variability tracking and maintenance) and activities related to individual products (e.g., change impact analysis and project progress monitoring) alike. Ideally, every  feature is first implemented on the platform level and only reused at the product level so that trace links from the platform are inherited by specific product instances to achieve full traceability. 
However, in practice, as reported by Company 2 and 3, this is rarely the case and product-specific features exist that make traceability on the product level necessary. It is a challenge to represent platform and product traceability in the TIM to make the traceability as visible and efficient as possible.


\item \textbf{No shared company-wide view on return on investment of trace links}
   Stakeholders defining a TIM need reasons for including certain relationship types in the TIM. 
   A common way to motivate a relationship type is to link it to a stakeholder need by eliciting the traceability needs of the different stakeholders in the company~\cite{mader2009getting,mader2013strategic}. For smaller systems, it is easier to understand which trace link types are needed since knowledge of how the different artifact types fit together is common, however for 
 complex systems, this elicitation leads to many suggested relationship types due to the large number of artifact types. They thus need to be filtered by the expected value they create, but it is difficult to predict the concrete value of certain relationship types. 
  Moreover, different link types have different value for different stakeholders. While this is a general traceability challenge, it is also TIM related since for complex systems it is unclear which combination of trace link types provide the most value. All the three companies in the workshop struggle with making decisions on which relationship types to include in the TIM. 
For example, in Company 2, there are more than 50 artifact types, which makes it difficult to decide which trace link types between these artifacts are actually valuable.

 \item\textbf{Varying requirements imposed by stakeholders with diverse needs}
A TIM is typically designed according to the stakeholder needs in the company. For large organisations, a single TIM needs to address the needs of many stakeholders. For instance, different stakeholders may not only require different link types but also require same link types but on different abstraction levels. 
This results in large and complex TIMs. The representation of such TIMs to stakeholders is difficult since different stakeholders need to derive different information from the TIM. For example, performing a system level safety analysis requires trace links on a higher abstraction level compared to low level safety analysis of specific functions. In practice, it is challenging to design a TIM that is flexible enough to support these different needs and still provide guidance to developers so that they create useful trace links using such a TIM.

\end{itemize}

\section{Large and distributed companies}
In this section, we describe challenges that are due to the companies being large and distributed but still having to coordinate and work together to create the same end product. 
\begin{itemize}
    \item \textbf{Difficult to define a TIM that supports mixed development processes and workflows}
     In large systems engineering companies, different disciplines (e.g., mechanical engineering, electrical engineering, and software engineering) need to be coordinated and integrated. Teams that focus on one of those disciplines tend to follow processes that are aligned with the way the parts of the product are developed. A mechanical engineer, e.g., is more likely to use a waterfall-ish process whereas a software engineer might work in an iterative-incremental fashion. Each process requires different kinds of artifacts, e.g., a product requirement document for the mechanical engineer and a backlog item for the software engineer. A TIM that is used in such an organisation needs to cater to all processes or workflows and therefore integrate all artifacts prescribed by all used processes, independent of discipline, thus contributing to the TIM's complexity. This challenge was reported by company 2 and 3.  
    \item \textbf{Difficult to align TIM 
    conformance checks  with project phases}
    Constraints defined in the TIM such as cardinality are used to check if links exist and are syntactically correct. For instance, if the TIM defines a one-to-one cardinality for a link from requirement to test, then this should be enforced to ensure that the link actually exists and the cardinality is adhered to. However, in practice, the enforcement of such 
conformance checks needs to be aligned with the current project phase or current phase within the sprint. For instance, the companies reported that 
enforcing that there is a link between a requirement and a test during requirements elicitation through the TIM is not helpful, but is helpful at a later stage when requirements are already gathered and need to be maintained. 
    This challenge becomes more pronounced when mixed development processes are used and there is no clear global definition of phases. Additionally, constraints on the TIM should not limit the agility of the different teams.  
\end{itemize}

\section{Long lasting products}
The products developed by the three companies have to be maintained over a long period of time. Thus the TIM and the traceability links in place also need to be maintained. This introduces three main challenges with respect to TIM management as organisations, development processes and system engineering techniques evolve. 
\begin{itemize}
    \item \textbf{Unclear strategies on how to evolve TIMs and trace models}
 All the companies acknowledged that once the TIM is defined and used, it is hard to change both on a technical and process level, since existing links conforming to the current TIM need to be migrated. Additionally, there are no strategies in place to analyse the existing traceability strategy in the companies to decide if the TIM needs to be updated. This leads to TIMs that are defined once and not updated, even though some of the relationships they capture have evolved.
    
    Currently, TIMs are changed in big bang approaches: when a new platform is created at Company 2, e.g., the TIM is changed as well. The old platform is maintained with the old TIM and the new platform will be maintained with the new TIM. However, the company now wants to move to a development strategy in which their platform will change less frequently, therefore the company needs a new strategy on how to continuously evolve the TIM.
    \item \textbf{Need to support artifact versions in the TIM}
In a typical software and systems engineering product line, various versions of an artifact co-exist and are used in products at the same time. Which version is used in which product needs to be tracked. In the automotive domain, e.g., regulations state that the manufacturer needs to trace which versions of the components are deployed in a specific vehicle. Trace links to specific artifact versions thus need to be established both at the platform level and product level (cf. Figure~\ref{fig:versioning_challenge}).

Company 2 and 3 reported that it is a challenge to define a TIM that explicitly supports versioning information for the artifacts connected by trace links. 

Company 1 reported creating traceability links from requirements and test cases to specific commits but having issues with tracking artifact versions when multiple artifact repositories are used.

\begin{figure}
\includegraphics[width=0.5\textwidth]{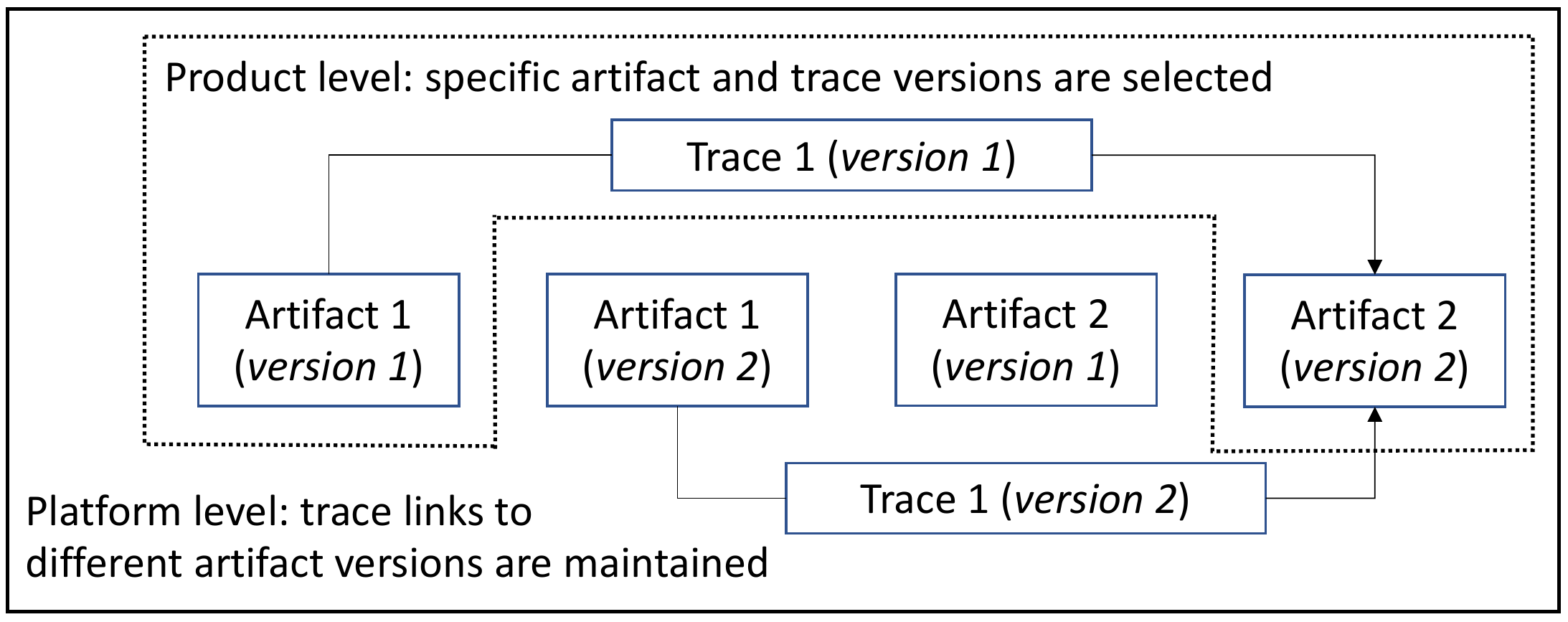}
\caption{Versioning of trace links in a product line environment.}
\label{fig:versioning_challenge}
\end{figure}
    \item \textbf{Complexity added when TIMs need to be adjusted to process maturity}
    The stability of a TIM reflects the stability of the development process. 
    For companies transitioning from one development process to another, TIM management is challenging. Product development continues in the midst of process changes and traceability still needs to be maintained. In such situations, it is necessary to manage and evolve multiple TIMs in parallel as different development teams have different schedules for adopting a new development process. For instance, Company 3 is transitioning from single product development to product-line development. In this transition, three TIMs are required: a TIM that supports a single product approach; TIM that supports a product-line approach; and a TIM that supports teams that are ``in between'', i.e., teams that have started to transition but are not yet fully transitioned. Such teams end up creating more traceability links than needed in either of the development paradigms, as they must trace to artifacts in the old and new process. Moreover, as teams migrate from one paradigm to another, trace link migration is difficult due to the challenge of evolving TIMs and trace models discussed before. 
    
\end{itemize}

\section{Refocusing TIM design research}
The existing guidelines on TIM management such as those from~\cite{ramesh2001toward}, \cite{mader2009getting} and~\cite{mader2013strategic} are a good starting point for practitioners. However, these guidelines do not cover all the challenges practitioners face when creating TIMs in large systems engineering companies. Particularly there is little support for understanding the implications of the development process aspect on the TIM, especially when mixed development processes are used. Our findings show that this is a problem in practice. 

For some of the challenges, e.g., TIMs for product lines, changeability of TIMs and TIM modularisation, research such as \cite{anquetil2010model}, \cite{paige2016evolving} and ~\cite{mader2010visual} exists respectively. However, this research only investigates the tooling aspect with restricting assumptions such as that all artifacts are persisted in certain formats and specific tools are used. This limits the transferability of such research to practice. 

Moreover, some of the guidelines are conflicting in practice. For instance, some literature suggests that the level of abstraction between two artifacts connected by a trace link should match~\cite{mader2009getting} while organisations that develop safety-critical products need to trace between artifacts on different levels of abstraction to be able to perform various safety analysis tasks required by safety standards. One of the solutions suggested is to trace to different levels of abstraction depending on the use of the trace links and have tool support to represent the links at different levels of abstraction~\cite{ramesh2001toward}. However, it is not clear how this can be handled in practice in a cost-efficient and effective manner. Additionally, existing TIM design guidelines are scattered in various publications making them hard to find, and there is no guidance on how practitioners can operationalize these guidelines.   

From our research, we believe that among many possible research directions, the following research is needed in the context of the challenges we discussed: 
\begin{itemize}
\item Research on managing complex TIMs such as TIM modularization, use of multiple TIMs in an organization and adoption of TIMs in a cost-efficient and effective manner.
\item Creation of guidelines for designing TIMs that support product-line platform related activities and product-specific activities, as well as how TIMs can support derivation of product-specific links from platform links. 
\item How to evolve TIMs and existing trace models such as using model transformations to transform existing links to match a new TIM, and how to evolve TIMs when processes change. 
\item Research on how organisations can design TIMs that support multiple development processes and workflows, e.g., by defining process-agnostic TIMs and providing guidelines on mapping the different artifact types and links to specific development processes.
\end{itemize}

While some of the above 
topics are not new to the requirements and traceability community, our aim is to 
explicitly point out these challenges to draw more attention to the practical complexity that exists in companies that deploy traceability. We acknowledge the usefulness of technical and tool-related research on traceability. However, from the perspective of managing TIMs, we believe solving many of the challenges requires a process-oriented research perspective. We therefore suggest more empirical studies specifically on TIM management focusing on large organisations developing complex product lines with long-lasting products. Only if research on traceability and TIM management acknowledges these practical challenges will it yield beneficial lessons for practitioners.



\section{Summary}
This paper reports on challenges that practitioners face when managing TIMs. The challenges were identified using a focus group with three companies from the automotive, industrial automation and telecommunication domain. The challenges are due to three drivers;  complexity of the products developed, the size and distributedness of the companies and the long lasting products that need to be maintained. Our study calls for more research on TIM management targeted to companies with such characteristics.



\section{ACKNOWLEDGMENTS}

We thank all company participants for their contributions to this paper. This work has been partially supported by Software Center Project 27 on RE for Large-Scale Agile System Development, the Wallenberg AI, Autonomous Systems and Software Program (WASP) funded by the Knut and Alice Wallenberg Foundation, the Sida/BRIGHT project 317 under the Makerere-Sida bilateral research program 2015-2020, and the ITEA 3 Call 6 project 17003 PANORAMA. 

\bibliographystyle{IEEEtran}
\bibliography{refs}

\begin{IEEEbiography}{Salome Maro} is a PhD candidate at
   \begin{wrapfigure}{l}{1in}
   \begin{center}
    \includegraphics[width=1in,height=1.25in,clip,keepaspectratio]{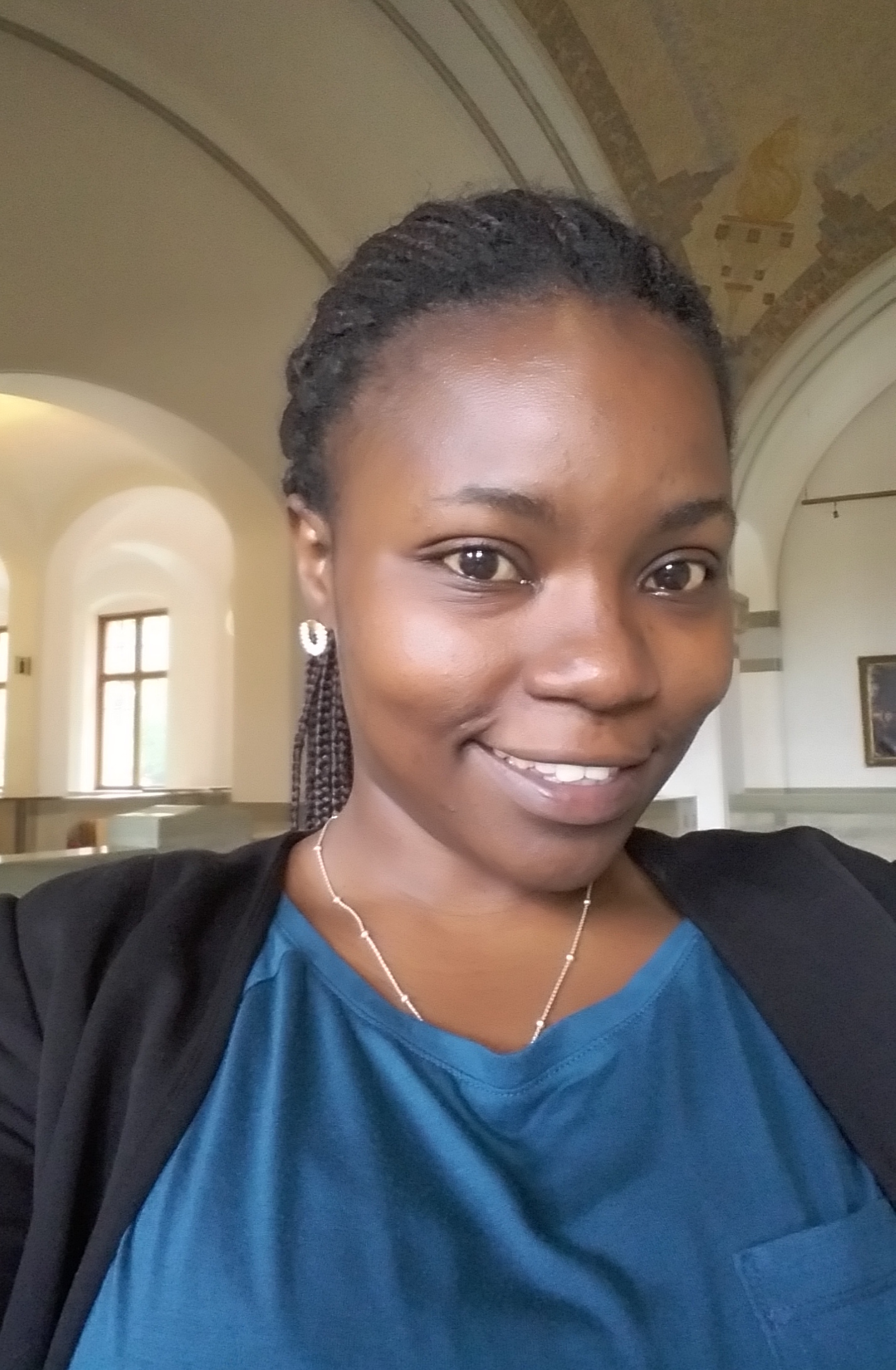}
   \end{center}
    \end{wrapfigure}
    Chalmers $|$ University of Gothenburg, Sweden.
       Her research is on software and systems traceability tools and processes in industry. She holds an M.Sc. degree in Software Enginering from University of Gothenburg. Contact her at salome.maro@cse.gu.se.
    \end{IEEEbiography}
    
\begin{IEEEbiography}{Jan-Philipp Stegh{\"o}fer} is an Associate Professor at 
   \begin{wrapfigure}{l}{1in}
   \begin{center}
    \includegraphics[width=1in,height=1.25in,clip,keepaspectratio]{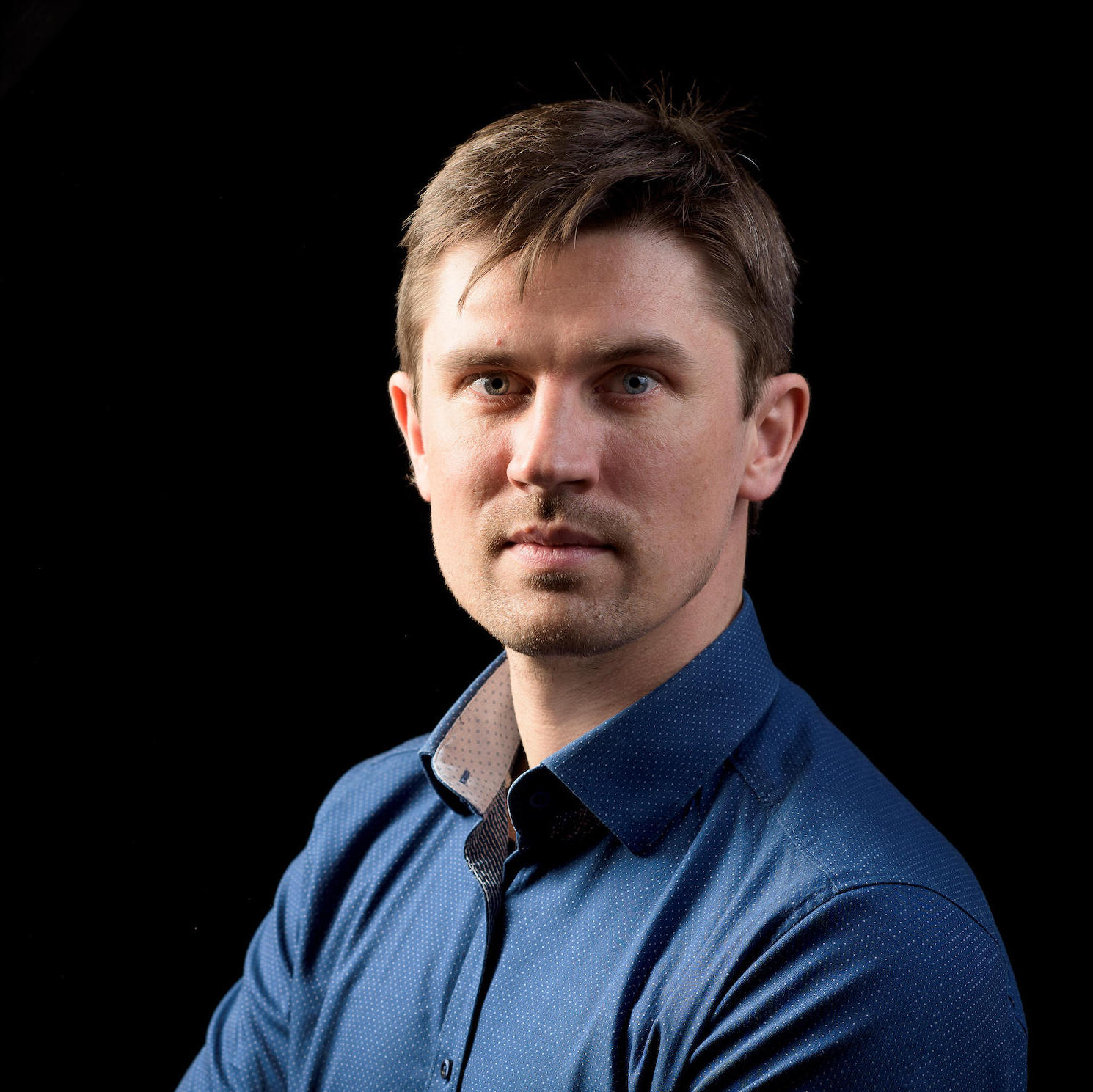}
   \end{center}
    \end{wrapfigure}
    Chalmers $|$ University of Gothenburg, Sweden.
    His research interests include software and systems traceability, large-scale and distributed agile development, software engineering education, and self-adaptive sytems.
    He holds a PhD from the University of Augsburg, Germany.
    Contact him at jan-philipp.steghofer@gu.se.
    \end{IEEEbiography}

\begin{IEEEbiography}{Eric Knauss} is an Associate Professor at 
   \begin{wrapfigure}{l}{1in}
   \begin{center}
    \includegraphics[width=1in,height=1.25in,clip,keepaspectratio]{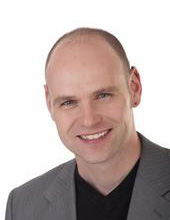}
   \end{center}
    \end{wrapfigure}
    Chalmers $|$ University of Gothenburg, Sweden.
    His research focuses on managing requirements and related knowledge in large-scale and distributed software projects.
    He holds a PhD from Leibniz Universit\"{a}t Hannover, Germany.
    He is member of program and organization committees of top conferences and reviewer for top journals.
    Contact him at eric.knauss@cse.gu.se.
    \end{IEEEbiography}
   
\begin{IEEEbiography}{Jennifer Horkoff} is an Associate Professor at
   \begin{wrapfigure}{l}{1in}
   \begin{center}
   \includegraphics[width=1in,height=1.25in,clip,keepaspectratio]{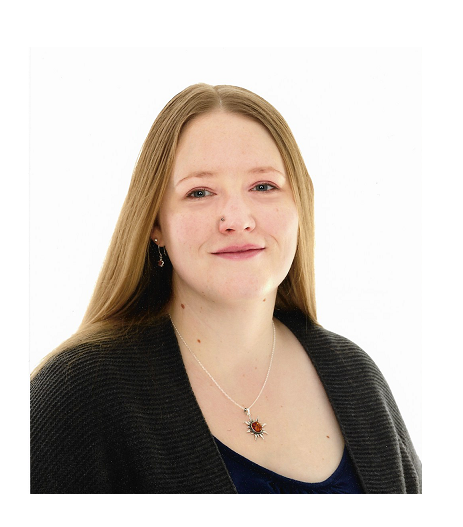}
   \end{center}
    \end{wrapfigure}
    Chalmers $|$ University of Gothenburg, Sweden.
  Her research focuses on requirements engineering, requirements modeling, quality and value modeling, model analysis and creativity. She holds a PhD from the University of Toronto. 
  Contact her at jennifer.horkoff@gu.se.
    \end{IEEEbiography}

\begin{IEEEbiography}{Rashidah Kasauli} holds a PhD from
 \begin{wrapfigure}{l}{1in}
   \begin{center}
    \includegraphics[width=1in,height=1.25in,clip,keepaspectratio]{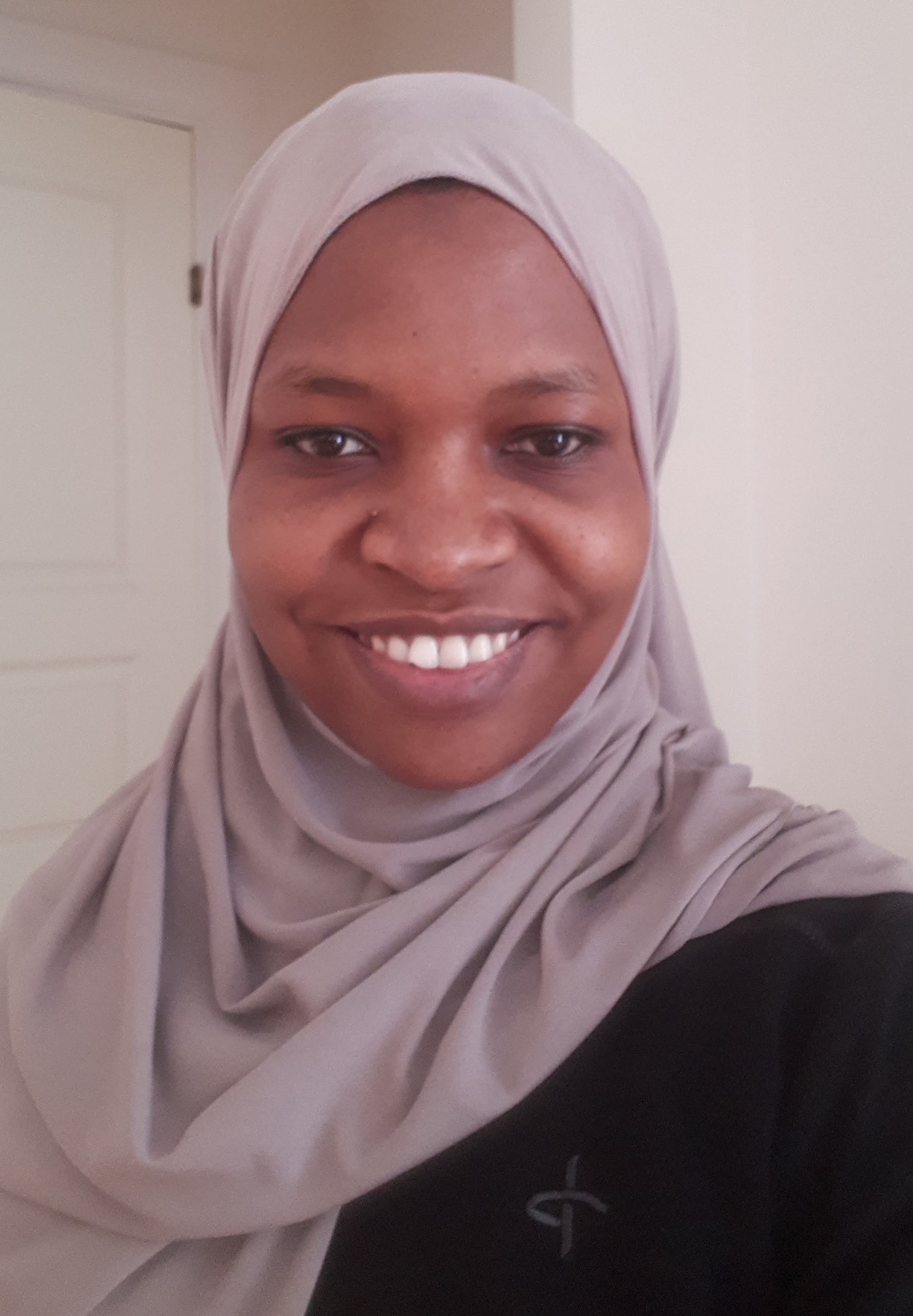}
   \end{center}
    \end{wrapfigure} 
    Chalmers $|$ University of Gothenburg, Sweden. 
    Her research focuses on requirements, safety-critical systems and large-scale agile development. 
    She holds an Msc in Data Communication and Software Engineering from Makerere University, Uganda. 
    Contact her at rashida@chalmers.se.
\end{IEEEbiography}

\begin{IEEEbiography}{Rebekka Wohlrab} holds a PhD from 
   \begin{wrapfigure}{l}{1in}
   \begin{center}
       \includegraphics[width=1in,height=1.25in,clip,keepaspectratio]{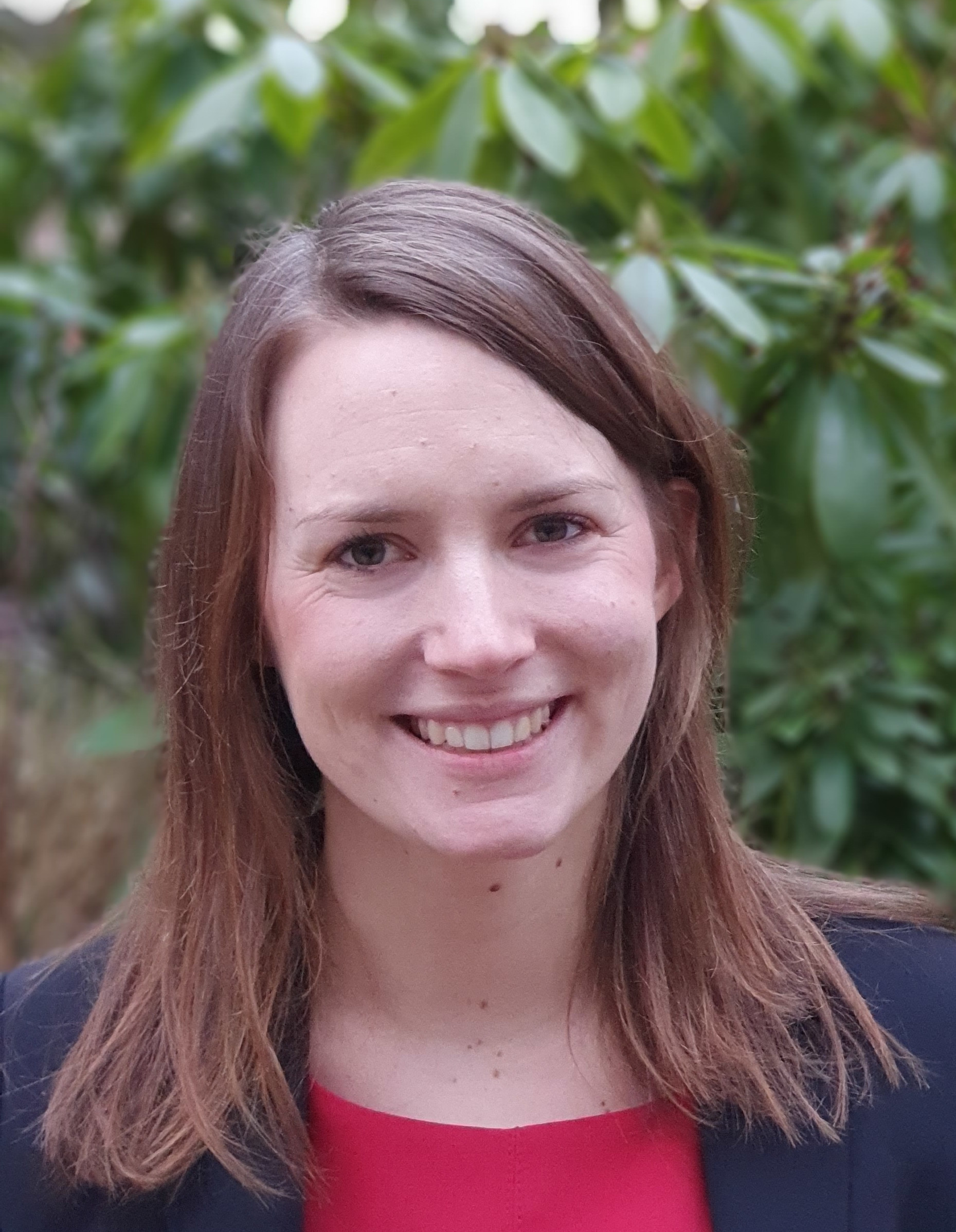}
   \end{center}
    \end{wrapfigure}
    Chalmers University of Technology, Sweden.
  Her research focuses on requirements engineering and software architecture in large-scale agile development. Contact her at wohlrab@chalmers.se.                                                                
 \end{IEEEbiography}
 
 \vspace{1cm}

  \begin{IEEEbiography}{Jesper Lysemose Korsgaard}  is Lead Systems Engineer
   \begin{wrapfigure}{l}{1in}
   \begin{center}
       \includegraphics[width=1in,height=1.25in,clip,keepaspectratio]{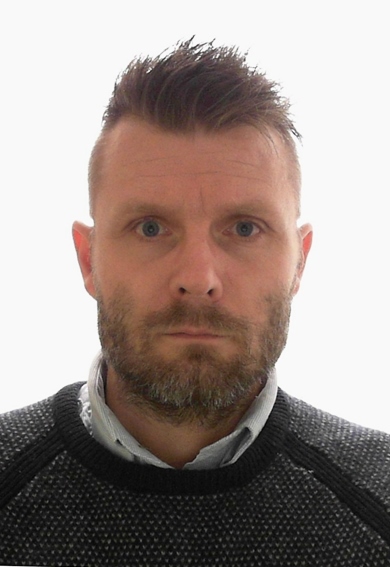}
   \end{center}
    \end{wrapfigure}
  also working with software requirements at GRUNDFOS Holding A/S.  Contact him at jlysemose@grundfos.com                                          
 \end{IEEEbiography}
 \vspace{2cm} 

\begin{IEEEbiography}{Florian Wartenberg} is a System Test Engineer 
   \begin{wrapfigure}{l}{1in}
   \begin{center}
       \includegraphics[width=1in,height=1.25in,clip,keepaspectratio]{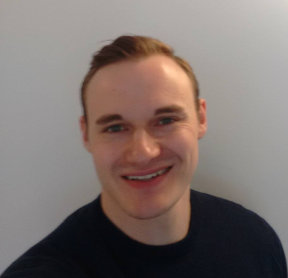}
   \end{center}
    \end{wrapfigure}
 at GRUNDFOS holding A/S. His interests focus on requirements, traceability, verification and validation.
He holds a diploma in Computer Science from Humboldt University of Berlin.
Contact him at fwartenberg@grundfos.com                                     
 \end{IEEEbiography}
 
 \begin{IEEEbiography}{Niels J\o{}rgen Str\o{}m}  has an SE-Zert Level B certification
   \begin{wrapfigure}{l}{1in}
   \begin{center}
       \includegraphics[width=1in,height=1.25in,clip,keepaspectratio]{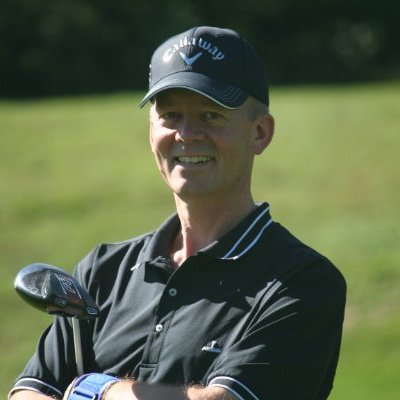}
   \end{center}
    \end{wrapfigure}
 and has a background as an electronics engineer and ISO assessor.
He is a Chief Systems Engineer at Grundfos Holding A/S and has been with the company for 25 years. He works with product lines, embedded systems, configuration management and process management. Contact him at njstroem@grundfos.com.                                
 \end{IEEEbiography}
 
  \begin{IEEEbiography}{Ruben Alexandersson} holds a Ph.D.
   \begin{wrapfigure}{l}{1in}
   \begin{center}
       \includegraphics[width=1in,height=1.25in,clip,keepaspectratio]{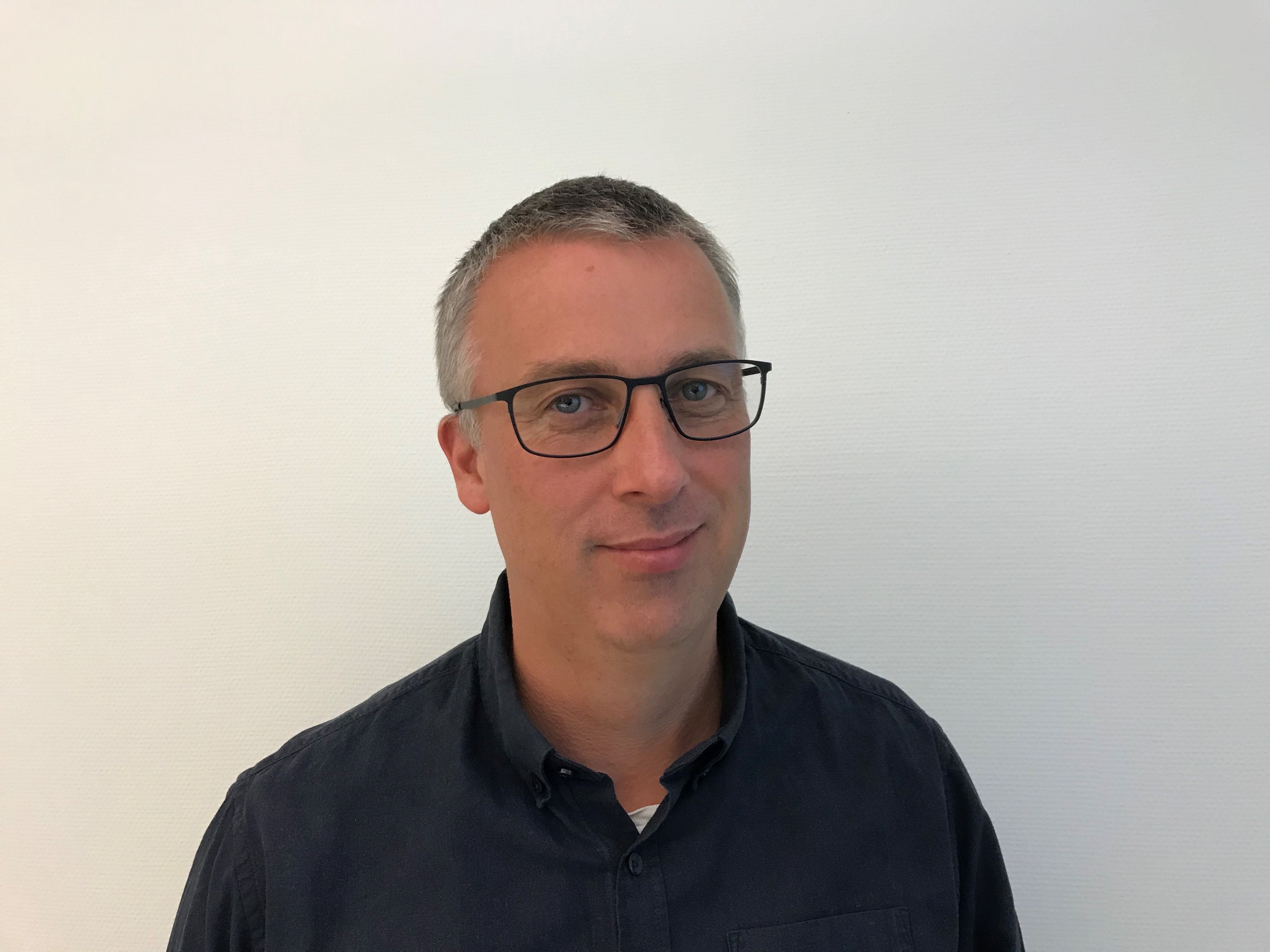}
   \end{center}
    \end{wrapfigure}
    from Chalmers University of Technology, and a background as teacher in Software Engineering. He is currently a Technical Specialist at Volvo Cars and has been with the company for ten years. Contact him at ruben.alexandersson@volvocars.com.                                                   
 \end{IEEEbiography}
 
\detailtexcount{tims-in-practice}

\end{document}

%% file: macro-editing.tex
\usepackage[normalem]{ulem} 
\usepackage{xcolor}

\newcommand{\del}[1]{\textcolor{red}{\sout{#1}}} 

\usepackage{ifthen}
\usepackage{amssymb}
\newboolean{showcomments}
\setboolean{showcomments}{true} 
\ifthenelse{\boolean{showcomments}}
  {\newcommand{\nb}[2]{
    \fcolorbox{gray}{yellow}{\bfseries\sffamily\scriptsize#1}
    {\sf\small$\blacktriangleright$\textit{#2}$\blacktriangleleft$}
   }
   
  }
  {\newcommand{\nb}[2]{}
   
  }

\newcommand\eric[1]{\nb{Eric}{#1}}

\newcommand\jp[1]{\nb{Jan-Philipp}{#1}}

\newcommand\salome[1]{\nb{Salome}{#1}}